\documentclass[preprint]{aastex} 
 
\def\spose#1{\hbox to 0pt{#1\hss}}
\def\lta{\mathrel{\spose{\lower 3pt\hbox{$\mathchar"218$}}
     \raise 2.0pt\hbox{$\mathchar"13C$}}}
\def\gta{\mathrel{\spose{\lower 3pt\hbox{$\mathchar"218$}}
     \raise 2.0pt\hbox{$\mathchar"13E$}}}
 
\shorttitle{ICPNe in Virgo} 
\shortauthors{Aguerri J.A.L., et al.} 
 
\received{2004 June 8}
\begin{document} 
 
 
\title{Intracluster stars in the Virgo cluster core} 
 
 
\author{J.A.L. Aguerri\altaffilmark{1,2}, O.E. Gerhard\altaffilmark{2}, 
M. Arnaboldi\altaffilmark{3}, N.R. Napolitano\altaffilmark{4}, 
N. Castro-Rodriguez\altaffilmark{1}, K.C. Freeman\altaffilmark{5}}  
 
\affil{$^1$ Instituto de Astro\'{\i}sica de Canarias. C/ V\'{\i}a L\'actea s/n, 38200 La Laguna, Spain. (jalfonso@iac.es; ncastro@iac.es)} 
\affil{$^2$ Astronomisches  Institut der Universitat Basel. Venusstrasse 7, CH-4102 Binningen, Switzerland. (ortwin.gerhard@unibas.ch)} 
\affil{$^3$ Osservatorio Astronomico di Pino Torinese, INAF, I-10025 Pino Torinese, Italy. (arnaboldi@to.astro.it)} 
\affil{$^4$ Kapteyn Astronomical Institute, Post Office Box 800, 9700 AV Groningen, Netherlands. (nicola@astro.rug.nl)} 
\affil{$^5$ Research School of Astronomy and Astrophysics, Mount Stromlo Observatory, Cotter Road, Weston Creek, ACT 2611, Australia. (kcf@mso.anu.edu.au)} 
 
 


\begin{abstract} 
  We have investigated the properties of the diffuse light in the
  Virgo cluster core region, based on the detection of intracluster
  planetary nebulae (PNe) in four fields.  We eliminate the bias from
  misclassified faint continuum objects, using improved Monte Carlo
  simulations, and the contaminations by high redshift Ly$\alpha$
  galaxies, using the Ly$\alpha$ luminosity function in blank
  fields. Recent spectroscopic observations confirm that our
  photometric PN samples are well-understood. We find that the diffuse
  stellar population in the Virgo core region is inhomogeneous on
  scales of $30'-90'$: there exist significant field-to-field
  variations in the number density of PNe and the inferred amount of
  intracluster light, with some empty fields, some fields dominated by
  extended Virgo galaxy halos, and some fields dominated by the true
  intracluster component. There is no clear trend with distance from
  M87. The mean surface luminosity density, its rms variation, and the
  mean surface brightness of diffuse light in our 4 fields are
  $\Sigma_B = 2.7 \times 10^{6}$ L$_{B\odot}$ arcmin$^{-2}$,
  $\mbox{rms} = 2.1 \times 10^{6}$ L$_{B\odot}$ arcmin$^{-2}$, and
  $\bar{\mu}_{B}=29.0$ mag arcsec$^{-2}$ respectively.  Our results
  indicate that the Virgo cluster is a dynamically young environment,
  and that the intracluster component is associated at least partially
  with local physical processes like galaxy interactions or
  harassment. We also argue, based on kinematic evidence, that the
  so-called 'over-luminous' PNe in the halo of M84 are dynamically
  associated with this galaxy, and must thus be brighter than and part
  of a different stellar population from the normal PN population in
  elliptical galaxies.
\end{abstract} 
 

 
\section{Introduction} \label{intro} 
 
The study of the intracluster light (ICL) began with Zwicky's (1951)
claimed discovery of an excess of light between galaxies in the Coma
cluster. Its low surface brightness ($\approx \mu_{B}>28$ mag
arcsec$^{-2}$) makes it difficult to study the ICL systematically
(Oemler 1973; Thuan \& Kormendy 1977; Bernstein et al.\ 1995; Gregg \&
West 1998; Gonzalez et al. 2000). A new approach was recently taken by
Zibetti et al.\ (2005), who stacked a large number of SDSS images to
reach deep surface brightness levels. In nearby galaxy clusters,
intracluster planetary nebulae (ICPNe) can be used as tracers of the
ICL; this has the advantage that detection of ICPNe is possible with
deep narrow-band images, and that also the ICPN radial velocities can
be measured to investigate the dynamics of the ICL component. ICPN
candidates have been identified in Virgo (Arnaboldi et al. 1996, 2002,
2003; Feldmeier et al. 1998, 2003a, 2004) and Fornax (Theuns \& Warren
1997), with significant numbers of ICPN velocities beginning to become
available (Arnaboldi et al.\ 2004).
 
The overall amount of ICL in galaxy clusters is still a matter of
debate.  However, there is now observational evidence that it may
depend on the physical parameters of clusters, with rich galaxy
clusters containing 20$\%$ or more of their stars in the intracluster
component (Gonzalez et al 2000; Gal-Yam et al. 2003), while the Virgo
cluster has a fraction of $\approx 10\%$ in ICL (Ferguson et al.\
1998; Durrell et al. 2002; Arnaboldi et al.\ 2002, 2003; Feldmeier et
al.\ 2004), and the fraction of detected intragroup light (IGL) is
1.3$\%$ in the M81 group (Feldmeier et al.\ 2003b) and less than
1.6$\%$ in the Leo I group (Castro-Rodr\'{\i}guez et al.\
2003). Recent hydrodynamical simulations of galaxy cluster formation
in a $\Lambda$CDM cosmology have corroborated this observational
evidence: in these simulated clusters, the fraction of ICL increases
from $\approx 10-20\%$ in clusters with $10^{14} M_{\odot}$ to up to
50$\%$ for very massive clusters with $10^{15} M_{\odot}$ (Murante et
al.\ 2004).
 
The mass fraction and physical properties of ICL and their dependence
on cluster mass will be related with the mechanisms by which the ICL
is formed.  Theoretical studies predict that if most of the ICL is
removed from galaxies due to their interaction with the galaxy cluster
potential or in fast encounters with other galaxies, the amount of the
ICL should be a function of the galaxy number density (Richstone \&
Malumuth 1983; Moore et al.\ 1996). The early theoretical studies
about the origin and evolution of the ICL suggested that it might
account for between 10$\%$ and 70$\%$ of the total cluster luminosity
(Richstone \& Malumuth 1983; Malumuth \& Richstone 1984; Miller 1983;
Merritt 1983, 1984). These studies were based on analytic estimates of
tidal stripping, or simulations of individual galaxies orbiting in a
smooth gravitational potential. Nowadays, cosmological simulations
allow us to study in detail the evolution of galaxies in cluster
environments (see for example Moore et al.\ 1996, Dubinski 1998,
Murante et al.\ 2004, Willman et al.\ 2004, Sommer-Larsen et al.\
2004).  Napolitano et al.\ (2003) investigated the ICL for a
Virgo-like cluster in one of these hierarchical simulations,
predicting that the ICL in such clusters should be unrelaxed in
velocity space and show significant substructures. The first radial
velocity measurements for a substantial sample of ICPNe (Arnaboldi et
al.\ 2004) have indeed shown significant field-to-field variations and
substructures, and spatial substructures have been observed in one
field in the ICPNe identified with [OIII] and H$\alpha$ (Okamura et
al.\ 2002).
 
The main goal of this paper is to estimate more reliably the amount of
ICL in the core region of the Virgo cluster, and to investigate the
homogeneity of the ICL distribution in this region.  Previously,
Ferguson et al.\ (1998) and Durrell et al.\ (2002) estimated the
amount of ICL from detecting excess red giant stars in two HST images,
while Arnaboldi et al.\ (2002, 2003) and Feldmeier et al.\ (2003a,
2004) obtained deep narrow-band images to detect ICPNe in several
Virgo fields, and then converted to IC luminosity using the bolometric
luminosity-specific PN number density for the evolved stellar
populations of the M31 bulge (Ciardullo et al.\ 1989a) or for Virgo
ellipticals (Jacoby et al.\ 1990).  This gave estimates between
10-20$\%$ for the fraction of ICL in different fields. It is not clear
at this stage whether this implies a genuine inhomogeneity, or whether
there are still systematic biases in the different determinations.
 
Here we study ICPNe in four wide field images in the Virgo cluster
core. Our observations and data reduction are described in
Section~2. Then we investigate possible biases in the detection of
ICPNe from narrow band surveys, and obtain a complete and homogeneous
sample of ICPNe in each of the four fields (Section~3). From these
data we determine the ICPN luminosity functions (Section~4) and the
surface densities of ICPNe and ICL in these four fields (Section~5).
Finally, we discuss the implications of our results in Section~6.
 
\section{Observations, Data Reduction and Sampled Fields}\label{obs} 
 
We wish to obtain in this work the average observational properties of
the diffuse light in the inner region of the Virgo cluster (we will
refer to this region as the core region). We have selected for this
purpose four fields, located within 80 arcmin ($\approx 350$ kpc) from
the cluster center at M87\footnote{Here and in what follows we compute
linear scales in Virgo with an adopted distance of 15 Mpc.}. Figure~1
shows a Digital Sky Survey (DSS) image of the Virgo cluster core
region, with the positions of the fields studied in this work
over-plotted. The fields were observed with different telescopes and
using different instrumentation, as follows.
 
In March 1999 we observed with the Wide Field Imager (WFI) on the
ESO/MPI 2.2-m telescope a field at $\alpha(J2000)=12:27:48$ and
$\delta(J2000)=+13:18:46$; here we will refer to this field as the
Core field.  The full image consists of a mosaic of eight 4k$\times$2k
CCD images, covering 34$'$$\times$34$'$ on the sky.  Each CCD has a
pixel size of $0''.238$, an average read-out noise of 4.5 ADU
pixel$^{-1}$, and a gain of 2.2 e$^{-}$ ADU$^{-1}$. We imaged this
field through an 80\AA\ wide filter centered at 5023\AA, the
wavelength of the [OIII] $\lambda 5007$\AA\ emission at the Virgo
cluster mean redshift.  In addition to this ``on-band'' filter, we
also imaged in the broad V band filter (the ``off-band'' filter).  We
took 8 individual images of 3000s for the narrow band filter and 8
images of 300s for the broad band filter.  Both broad and narrow band
images were obtained under photometric conditions, and the seeing in
the final combined images was $1''.2$.
 
In February 2001 we acquired another field with the Wide Field Camera 
(WFC) at the 2.5m Isaac Newton Telescope at the Roque de los Muchachos 
Observatory (La Palma). The central coordinates of this field are: 
$\alpha (J2000)= 12:25:32$ and $\delta (J2000)= +12:14:39$. The WFC 
detector is made of 4 CCDs with a total field of view of 34$'$ 
$\times$ 34 $'$. The pixel scale is $0''.333$, and the mean read-out 
noise and gain of the detectors are 6.1 ADU pixel$^{-1}$ and 2.8 
e$^{-}$ ADU$^{-1}$, respectively. We will refer hereafter to this 
field as the La Palma core (LPC) field. The narrow band filter used 
in these observations was a 60\AA-wide filter centered at 5027 
\AA.  The broad band filter was centered on the B band (4407\AA) 
and was 1022\AA\ wide. The total exposure time was 27000 secs for 
the narrow band image and 5400 secs for the broad band image. The 
seeing in the final combined images was $1''.5$ in both filters. 
 
We will also analyze in the present paper the ICPNe located in one of 
the fields with photometry from Feldmeier et al.\ (1998), at 
$\alpha(J2000)=12:30:39$ and $\delta(J2000)=+12:38:10$, and the ICPNe 
in a field one degree north of the Virgo core with photometry from 
Arnaboldi et al.\ (2002), hereafter Paper I. These will subsequently be 
referred to as the FCJ and RCN1 fields, respectively. The RCN1 field 
is located outside the Virgo core, but will be very useful for studying 
the ICPN density profile. 
 
Our sample of fields in the Virgo cluster core is completed with a 
field located at $\alpha(J2000)=12:25:47$ and 
$\delta(J2000)=+12:43:58$; we will refer to this field as the Subaru 
core (SUB) field (see Arnaboldi et al. 2003, hereafter Paper II). 
This field was imaged with the Suprime-Cam 10k$\times$8k mosaic 
camera, at the prime focus of the Subaru 8.2m telescope.  Images were 
acquired through two narrow-band filters, corresponding to [OIII] and 
H$\alpha$ emission at the redshift of Virgo cluster, and two broadband 
filters, the standard V and R. 
 
The data reduction and calibration of the two new fields presented in 
this paper (Core and LPC) were carried out as for the RCN1 field in 
Paper I. The data reduction was performed using the MSCRED package in 
IRAF. For a detailed discussion of the mosaic data reduction we refer 
the reader to Alcal\'a et al. (2002, 2004). Calibrations were obtained 
using several Landolt fields for the broad band filters, and 
spectrophotometric stars for the narrow band filters. Fluxes were then 
normalized to the AB magnitude system, following Theuns \& Warren 
(1997).  A detailed description of the calibration steps and the 
relation between AB magnitudes and the ``$m(5007)$'' [OIII] magnitude 
introduced by Jacoby (1989) are given in Paper I for the FCJ and RCN1
fields, and in Paper II for the SUB field.  The relation between AB 
magnitude and $m(5007)$ for the observations of the fields analyzed in 
the present paper are given in Table~1. In this paper we will use the 
notation of $m_{n}$ and $m_{b}$ to refer to the narrow-band and 
broadband magnitudes of the objects in the AB system. We will use 
$m(5007)$ to refer to the [OIII] magnitude introduced by Jacoby 
(1989). 
 
Table~1 gives the main observational characteristics of the fields
analyzed in this paper.

\section{Photometry: catalog extraction and validation} 
 
We used the automatic procedure developed and validated in Paper I for
performing the photometry and identification of emission line objects
in our mosaic images, in a homogeneous fashion (see Paper I for a full
description of the procedure). This procedure was applied to all
fields in order to obtain the ICPN photometric candidates.  The
automatic extraction procedure starts with measuring the photometry of
all objects in the images; this is done using SExtractor (Bertin \&
Arnouts 1996). All objects are plotted in a color magnitude diagram
(CMD) m$_{n}$-m$_{b}$ vs. m$_{n}$, and are classified according to
their positions in this diagram.  The most reliable ICPNe photometric
candidates are point-like sources with no detected continuum emission
and observed EW\footnote{We have computed the observed equivalent
width ($EW_{obs}$) using the expression given by Teplitz et al (2000):
$EW_{obs} \approx \Delta \lambda_{nb}(10^{0.4\Delta m}-1)$, where
$\Delta \lambda_{nb}$ is the width of the narrow-band filters in \AA,
and $\Delta m=m_{n}-m_{b}$ is the color of the object, with $m_{n}$ and
$m_{b}$ the narrow-band and broadband magnitudes, respectively.}
greater than 100\AA, after convolution with the photometric errors as
a function of magnitude\footnote{The fact that the selection procedure
is based on the observed EW greatly reduces the contamination of the
ICPNe candidate sample by [O II] emitters at $z=0.35$ whose emission
lines would also fall into the narrow band filter bandpasses used for
the Virgo photometry. Colless et al. (1990), Hammer et al. (1997) and
Hogg et al. (1998) found no [O II] emitters at $z=0.35$ with observed
EW greater than 95\AA.} (see Paper I for more details about the
selection criteria). Figure~2 shows the CMD for the Core, FCJ and LPC
fields.  The final photometric ICPN catalogs, following the selection
procedure explained, contain 117, 36 and 14 objects for the Core, FCJ
and LPC fields, respectively (see Table~2).  The ICPNe in the SUB
field were selected from the two-color diagram $[{\rm
OIII}]-H_{\alpha}$ versus $[{\rm OIII}]-(V+R)$, calibrated using
planetaries in M84 (see Paper II for the full description of this
method). The number of ICPNe detected in the SUB field was 36, and in
RCN1 the number of candidates was 75 (Paper I).
 
The field-to-field variations observed in the number of ICPN
photometric candidates are due in part to the different limiting
magnitudes and areas covered by our fields. Limiting magnitudes were
obtained following the simulation procedure described in Paper I. We
randomly distribute point-like objects in our final co-added [OIII]
images, assuming an exponential luminosity function (LF).  The
photometry of these objects was measured with SExtractor using the
same parameters as for the real sources. We define the limiting
magnitude for one of our images as the faintest magnitude at which
half of the input simulated sample is still retrieved from the
image. Table~1 shows the narrow-band and off-band limiting magnitudes
for all fields. The number of ICPN candidates down to the respective
[OIII] limiting magnitude is 77, 20, 14 and 36 for the Core, FCJ, LPC
and SUB fields, and 55 for RCN1 (see Table~2).
 
The photometric selection methods can be spectroscopically validated
by taking individual spectra of the selected candidates. This is
discussed in more detail in Section~3.4 below.  In the rest of this
section we will study the possible sources of contaminants for the
Core, FCJ, LPC, and RCN1 fields, and we will statistically subtract
the number of contaminants in the photometric samples.  This will
provide more secure ICPNe sample sizes, as compared to Paper I.
 
We will not study the contaminants in the sample of emission line 
objects obtained in Paper II for the SUB field. This sample was 
selected from the detection of emission in two narrow bands ([OIII] and 
H$\alpha$).  Spectroscopic confirmation was carried out in this field, 
for a subsample of 10 objects, at the 3.5m Telescopio Nazionale 
Galileo with the DOLORES spectrograph, in multi object spectroscopy 
(see Paper II). The result was that 8/10 photometrically selected 
objects were spectroscopically confirmed as emission line objects. The 
other 2 were not detected, probably due to astrometric problems and/or 
incorrect positioning of the slits (see Paper II).

\subsection{Contamination by faint continuum objects} 
 
It is possible that the ICPN samples in the Core, FCJ, LPC, and RCN1 
fields may be contaminated by mis-classified faint continuum objects, 
because of the selection based on a threshold in their [OIII] fluxes. 
Because of the photometric errors in their [OIII] fluxes, some objects 
are assigned a brighter flux than their real flux, others a fainter 
flux.  Because also their LF rises towards faint magnitudes, a significant 
number of objects will have measured [OIII] magnitude brighter than 
$m_{\rm lim}[{\rm OIII}]$ (or narrow-band AB $m_{{\rm lim},n}$), even 
though in reality they are fainter than the limiting magnitude.  If in 
addition their fluxes in the off-band image are below the limiting 
magnitude of that image, these objects will appear in the region of 
the CMD populated by the selected ICPN candidates, and they will be 
counted as ICPN candidates even though they are continuum objects 
(faint stars).  We will call this {\it the spill-over effect} from 
faint stars. 
 
Near the $m_{{\rm lim},n}$ limiting magnitude, the number of 
spill-over stars will be negligible if and only if the off-band image is 
deep enough for detecting the weak continuum flux of these faint 
objects.  This requires that the off-band image has an AB limiting 
magnitude of at least  
\begin{equation} 
m_{{\rm lim},b}\simeq m_{{\rm lim},n} + 3 \times <{\rm rms}> 
\end{equation} 
where $<{\rm rms}>$ is the mean photometric error of objects with
magnitude equal to $m_{{\rm lim},n}$.  This assumes that the AB-color
of the stars between off-band and on-band is zero, and leaves some
margin for the fact that the photometric errors will increase towards
fainter magnitudes beyond $m_{{\rm lim},n}$.  The $<{\rm rms}>$ values
for our fields are 0.27, 0.24, 0.19, and 0.23 for the Core, FCJ, LPC,
and RCN1 fields, respectively. Then, the corresponding off-band images
should be 0.81, 0.72, 0.57, and 0.69 magnitudes deeper than
the respective on-band images. Table~1 shows the actual limiting
magnitudes for the on and off-band images of our fields. We can see
that the off-band image in the Core field is only as deep as the Core
on-band image, suggesting a significant spill-over contamination,
while the RCN1, FCJ, and LPC fields have off-band images which are
deeper than their on-band images by 0.55, 0.6 and 0.9 magnitudes,
respectively.  In these fields, we can expect that the spill-over
effect will be less important.

We have performed simulations in order to quantify the spill-over 
effect in the Core, FCJ, LPC, and RCN1 fields. First, we selected from the 
photometric catalog of each field as continuum objects those with 
color $m_{n}-m_{b}=[-0.1,0.1]$. Then we fitted an exponential LF to 
these objects. For the simulations we distributed randomly on the 
scientific images several thousand point-like continuum sources, with 
the same LF as the real continuum objects but extrapolated down to 
$m_{n}=28$.  This faint-end magnitude in the simulated continuum 
sources is $\approx 3$ magnitudes fainter than the $m_{lim,n}$ of the 
images. The simulated objects were then recovered from the images, 
using the same technique and criteria to measure their photometry as 
for all sources in the real images. Based on the catalog obtained, we 
finally constructed the CMD for the recovered simulated objects, as 
shown in Figure~3.  To quantify the spill-over effect for each field, 
we computed the number of simulated objects that are brighter than the 
corresponding $m_{lim,n}$ and are located in the region of the ICPN 
photometric candidates.  After scaling the simulations to the same 
number of observed continuum objects, the final number of spill-over 
contaminants brighter than the $m_n$ (or [OIII]) limiting magnitude is 
45, 4, 2, and 16 for the Core, FCJ, LPC, and RCN1 fields, corresponding to 
58\%, 20\%, 14\%, and 29\% of the ICPN candidates brighter than 
$m_{{\rm lim},n}$ in these fields, respectively (see also Table~2).

\subsection{Missing ICPNe in the photometric samples} 
 
Due to the photometric errors, an emission line object may have 
smaller measured EW or fainter $m_n$-magnitude than it would have 
intrinsically. This means that we can lose in our catalogs some of the 
emission line objects with intrinsic $EW_{obs}>100$\AA\ above the 
magnitude limit. To investigate this possible source of error in our 
final ICPN photometric samples, we have quantified the number of such 
objects with simulations. We randomly distributed point-like objects 
on the narrow-band images following an exponential LF similar to the 
LF of the PNe in M87 (Ciardullo et al.\ 1998).  No broadband emissions 
were assigned to these objects.  We measured the photometry of them 
just as for the real sources, and studied their distribution in the 
CMD. We found that 5\%, 10\%, 1\%, and 3\% of these objects are lost 
in our catalogs for the Core, FCJ, LPC, and RCN1 fields, 
respectively (see Table~2). The most important factor that determines 
these fractions is again the limiting magnitude of the broad-band 
image; in deeper V-band images, the smaller rms noise makes it less 
likely that an object without continuum can be assigned 
$EW_{obs}<100$\AA. 
 
\subsection{Contamination by background galaxies\label{lyalpha}}  
 
The photometric samples of ICPNe can also be contaminated by emission
line background galaxies. This is because for [OII] starburst galaxies
at $z\simeq0.35$ and Ly$\alpha$ galaxies at $z\simeq3.1$ their strong
emission lines fall into our narrow band filter width.  As pointed out
before, the threshold in EW implied by our selection criteria ensures
that the selected ICPN photometric samples are nearly free of [OII]
contaminants (as confirmed also by the available spectroscopic
follow-up observations). However, some Ly$\alpha$ galaxies at $z\simeq
3.1$ can contaminate our ICPN photometric samples.  Spectroscopic
follow-up observations of ICPNe indeed found that a fraction of the
ICPN candidates were Ly$\alpha$ objects (Freeman et al.\ 2000,
Kudritzki et al.\ 2000, Arnaboldi et al.\ 2004).
 
Recently, we imaged an area in the Leo group (Castro-Rodr\'{\i}guez et 
al.\ 2003), and extracted emission line objects using the same 
selection criteria as for the ICPN photometric candidates in the Core, 
FCJ, LPC, and RCN1 fields. The LF of the emission-line objects 
selected in the Leo image had a bright cut-off $\approx$ 1.2 mag 
fainter than for the LF of the PNe associated with the elliptical 
galaxies in the Leo group. The most plausible explanation for this 
result is that the selected emission-line objects in the image of the 
Leo group are Ly$\alpha$ background galaxies; two of these were indeed 
spectroscopically confirmed as Ly$\alpha$ galaxies (see 
Castro-Rodr\'{\i}guez et al.\ 2003).  This gives us the opportunity to 
use the Leo image as a blank field to evaluate the contamination of 
background galaxies in our ICPN photometric catalogs\footnote{We have 
  also simulated the spill-over effect in the Leo images. We found 
  that only 6\% of the emission line sample in this field should be 
  due to the spill-over from faint stars.}.  Taking into account the 
differences in the surveyed areas and the narrow-band filter widths 
for the Leo and the other fields, we can compute the number of 
expected Ly$\alpha$ contaminants brighter than the [OIII] limiting 
magnitude in each field, based on the surface density of emission line 
sources in the Leo field.  The result is 20, 2, 22, and 3 expected 
Ly$\alpha$ contaminants in the Core, FCJ, LPC, and RCN1 fields (see 
Table~2). 
 
However, there appear to be intrinsic field-to-field variations in the 
surface density and luminosity function of Ly$\alpha$ emitters. 
Figure~4 shows the LF of the emission objects from the Leo field, the 
spectroscopically confirmed Ly$\alpha$ sample from Kudritzki et al.\ 
(2000), and the background sources from the blank field survey of 
Ciardullo et al.\ (2002a).  It is important to notice that the Leo and 
Ciardullo et al.\ (2002a) fields have similar off-band limiting 
magnitudes. But the brightest emission objects in the Ciardullo et 
al.\ field is 0.6 mag brighter than in the Leo field, and is 
consistent with the brightest Ly$\alpha$ emitters in the Kudritzki et 
al.\ (2000) sample. Using an average Ly$\alpha$ luminosity function 
obtained from the Ciardullo et al.\ and Kudritzki et al.\ samples, 
we predict the number of Ly$\alpha$ contaminants in the Core, FCJ, 
LPC, and RCN1 to be 26, 4, 16, and 26.  
 
Thus there is no unique way to estimate the number of high-redshift
emission galaxies in our photometric samples, and we have to rely on
spectra as much as possible. Fortunately, the predicted number of
Ly$\alpha$ contaminants is sensitive to the assumed luminosity
function only in the RCN1 field.  As discussed further in the next
subsection, there are very few Ly$\alpha$ emitters in the
spectroscopically confirmed emission samples of Arnaboldi et al.\
(2004) in the Core, FCJ, and SUB fields, down to
$m(5007)=27.2$.  Based on this result, we will use the Leo blank field
luminosity function to estimate the Ly$\alpha$ fraction in these
fields in Table~2 and what follows.
 
For the LPC field, Figure~4 shows that the bright cut-off of the
emission sources is similar to the bright end of the LF in the samples
of Kudritzki et al.\ (2000) and Ciardullo et al.\ (2002a), while none
of these candidates is as bright as the brightest PNe observed in
M87. Furthermore, the number of candidates in the LPC field is small,
and is in fact smaller than the estimated number of Ly$\alpha$
galaxies expected in this field for either of the two Ly$\alpha$
luminosity functions, down to our magnitude limit. These facts
together suggest strongly that all our candidates in this field,
including the bright objects, are Ly$\alpha$ emitters. We therefore
assume in Table~2 and the subsequent discussion that the number of PNe
in the LPC field is $<1$.
 
Finally, in the RCN1 field, we have no grounds to prefer one
Ly$\alpha$ luminosity function over the other, and will thus give the
number of expected PNe in this field based on both luminosity
functions.
 
\subsection{Comparison with spectroscopic results} 
 
Arnaboldi et al.\ (2004) carried out the spectroscopic follow-up of our 
selected ICPN samples in the FCJ and Core fields, and of the ICPN candidates 
from Okamura et al.\ (2002) in the SUB field. Spectroscopic 
observations were done in service mode with the FLAMES spectrograph at 
UT2 on VLT, and exposure times were such as to ensure the detection of
ICPN candidates with $m(5007) = 27.2$, equivalent to $4.2\times10^{-17}$
erg cm$^{-2}$ sec$^{-1}$ total flux in the line,  with S/N$\simeq 5$. 
 
A total of 70 fibers were allocated to candidates from the above
samples with $m(5007) \leq 27.2$: 18/34/18 FLAMES fibers were
allocated to sources brighter or equal to 27.2 in the FCJ/Core/SUB
fields, respectively. Arnaboldi et al.\ (2004) detected a total of
15/12/13 sharp line emitters, and 0/2/0 L$\alpha$ emitters which show
one resolved asymmetric line in these FLAMES pointings. The fraction
of confirmed spectra with both components of the [OIII] doublet
detected was 67\%/41\%/18\%. {Given the observing conditions and
S/N ratio achieved, these fractions are consistent with the assumption
that all of the sharp line emitters are indeed PNe.  Thus, for [OIII]
fluxes such that $m(5007) \leq 27.2$, the contamination by background
galaxies is small.
 
The remaining spectra in the Arnaboldi et al.\ (2004) observations did
non show any spectral features in the $\lambda$ range covered by
FLAMES. As we have seen in the previous subsections, one should indeed
expect a fraction of candidates in the candidate samples to be
contaminants due to the spill-over effect, i.e. faint continuum stars
erroneously classified as ICPNe because of a shallow off-band
image. Thus, the predicted spectroscopic confirmation rates, given as
the fraction of true ICPNe in these candidate samples, varied strongly
from field to field, despite similar $m_{lim}(5007)$.  The
spectroscopic follow-up, as well as earlier spectroscopic observations
by Kudritzki et al. (2000) and the comparison with deeper narrow band
imaging in [OIII] and H$\alpha$ (Paper II), provide direct evidence
for contamination by faint stars in narrow band [OIII] ICPN
surveys. Contamination from faint stars is also a possible explanation
for the somewhat low spectroscopic confirmation rate for the PNe
samples in ellipticals (Arnaboldi et al.\ 1996, 1998).

Arnaboldi et al.\ (2004) compared the expected spectroscopic
confirmation rates based on the simulations in the present paper with
those determined with FLAMES. They are in close agreement, showing
that the photometric samples are now well understood. A high
confirmation rate (small contamination) can be achieved when the
off-band image is sufficiently deep.

\section{ICPNe Luminosity Function}

\subsection{The necessity of deep off-band images} 

Figure~5 shows the LFs of the ICPN photometric candidates (full
points) in all fields but LPC. We have also over-plotted the LFs of
miss-classified stars (asterisks). A comparison with Figure~4 shows
that the LF of the Ly$\alpha$ galaxies has a fainter cut-off than the
LF of the photometric ICPN candidates in these fields. Thus the
brightest bins of the IC PNLF are free of Ly$\alpha$ contaminants.
There exist brighter Ly$\alpha$ galaxies, but these often have a
measurable continuum (Kudritzki et al.\ 2000).  The brightest LF bins
are also free of miss-classified stars in most fields; however, in the
Core field the IC PNLF is affected by miss-classified stars at all
magnitudes. This demonstrates again the importance of obtaining deep
broad-band images.  Figure~5 also contains the LF of the ICPN
candidates from the SUB field, which is free of contaminants in all
bins.
 
We have fitted the standard PNLF to the three brightest magnitude bins
(not affected by contaminants) of the IC PNLF in the FCJ and RCN1
fields. The six brightest bins were considered in the fit of the IC
PNLF in the SUB field. In both cases, the LF was first convolved with
the photometric errors.  The fitted empirical exponential LF is given
by
\begin{equation} 
N(m)=c_{1}e^{c_{2}M}[1-e^{3(M^{*}-M)}] \label{PNLF} 
\end{equation}
where $c_{1}$ is a positive normalization constant to be determined,
$c_{2}=0.307$, and we adopt a cutoff $M^{*}(5007)= -4.51$ (Ciardullo
et al.  1989b, Ciardullo et al.\ 2002b). The resulting apparent
magnitudes of the bright cutoffs (m$^{*}$) obtained from the fits are
$25.9\pm 0.2$, $26.0\pm 0.3$, and $26.3\pm0.15$ for the RCN1, FCJ and
SUB samples.  As for reference, we report here the m$^{*}$ value
fitted to the PNLF from the inner region of M87, which amounts to
$26.37$.

\subsection{The nature of over-luminous PNe and the bright cutoff of the LF} 
 
The bright cutoff of the PNLF has been extensively used as a
distance indicator in extragalactic astronomy (Ciardullo et
al. 2002b). An implicit assumption of the PNLF method is that all the
PNe that make up the observed luminosity function are at the same
distance. In a Virgo-like cluster, its finite depth can distort the
PNLF, as the ICPNe at shorter distances can contribute to the observed
LF and produce a brighter cutoff, as discussed in Ciardullo et
al. (1998), Arnaboldi et al.\ (2002), and Feldmeier et al.\ (2004).

However, the bright cutoff can be also affected by the age (Marigo et
al.\ 2004) or metallicity (Dopita et al.\ 1992) of the stellar
population.  E.g., the dependence of $M^{*}$ with metallicity, $\Delta
M^{*}=0.928[O/H]^{2}+0.225[O/H]+0.014$ (Dopita et al.\ 1992, Ciardullo
et al.\ 2002b) can produce a change of the cutoff magnitude of a
similar order to the reported line-of sight (LOS) distortions.  The
analysis of our ICPN LFs combined with the spectroscopic results of
Arnaboldi et al.\ (2004) can clarify the relative importance of these
effects.

The bright cutoff of our ICPN LF in the FCJ field is $0.37$ magnitudes
brighter than the bright cutoff of the PNLF of M87 (Ciardullo et al.\
1998).  In Paper I we considered that this (substantial) difference
would most plausibly be explained if the ICPNe in the FCJ field are
located in front of M87.  Assuming a distance of 15 Mpc for M87, the
brightest of the intracluster PNe in the FCJ field would be located
2.3$\pm1.5$ Mpc in front of M87, indicating a substantial LOS depth of
the Virgo cluster, consistent with other measurements.

The spectroscopic follow-up in FCJ has shown a velocity distribution
which is dominated by a narrow peak related to the halo of M87
(Arnaboldi et al.\ 2004). In addition to this peak, there are three
kinematic outliers, which are all also unusually bright, with fluxes
in the bright falloff of the M87 PNLF, so that they shift the FCJ LF
to a brighter cutoff. The correlation between their discrepant radial
velocities and bright $m(5007)$ support the hypothesis that these are
ICPNe free flying in Virgo intra-cluster space and not gravitationally
bound to M87. LOS effects are therefore a viable explanation for the
brighter cutoff in this field.

No spectroscopic follow-up is avaliable for the RCN1 sample yet, but
the absence of bright galaxies and a very similar PNLF to that in FCJ
suggest that LOS effects may be the most likely explanation for the
bright cutoff in this field also, as already discussed in Paper I.

The PNLF for the ICPN sample in the SUB field differs from RCN1 and
FCJ: 5 ICPNe appear to be up to 0.3 mag brighter than the remaining
smoother, more regular, distribution, similar to those measured for
ICPNe in FCJ and RCN1 (see Fig.~5).  Can the over-luminous ICPNe in
SUB be on the near side of the cluster?  The spectroscopic follow-up
of Arnaboldi et al.\ (2004) contained three of these objects and
confirmed them as PNe. Their radial velocities are consistent with the
dynamics of M84, however. The kinematic observations make it unlikely
that these objects are 2 Mpc in front of M84; most likely they are at
the same position as M84 within the Virgo cluster, and LOS effects are
unimportant.  Furthermore, the PNLF fit to only the remaining,
smoother distribution gives a distance modulus of 30.95$\pm$0.15; from
the LF of PNe associated directly with M84, the distance modulus of
M84 is 30.98 (Jacoby et al.\ 1990).  Thus, within the errors, M84 and
the bulk of the IC stellar population in the SUB field are at the same
distance.

The other alternative for the over-luminous PNe in SUB is that they
came from a different stellar population, which could be either
younger or of different metallicity from the main M84 stellar
population (see Marigo et al.\ 2004). Perhaps they are associated with
the infall of some dwarf galaxies into M84. Furthermore the PNe in
this field show a non-uniform spatial distribution.  Most of them are
located in a loop near M84 (see Okamura et al.\ 2002). This also
supports the scenario that they are associated with some harassment or
infall process involving M84.

\subsection{The relative distances of M87 and M84} 

Neilsen \& Tsvetanov (2000) and West \& Blakeslee (2000) have used
surface brightness fluctuation (SBF) distances to consider the
three-dimensional structure of the Virgo cluster ellipticals and found
that most of these are arranged in a nearly co-linear structure
extending $\pm (2-3)$ Mpc from M87.  The M84/M86 subgroup is falling
into the Virgo cluster from behind, and is itself on the far side of
the mean Virgo cluster.

We have tested the relative distance of these sub-units in Virgo based
on the most recent PNe data sets obtained for these galaxies, the
Ciardullo et al.\ (1998) sample for M87 and the Paper II data for M84.
We have run a Kolmogorov-Smirnoff (KS) test of the PNLFs associated
with M84 and the inner $5'$ region of M87, without the over-luminous
PNe.  The KS test cannot exclude that the two LFs came from the same
distribution function. Thus, the PNLF method based on recent PN
datasets places M84 and M87 at similar distances.

\section{Surface density and fraction of IC light in different fields} 
 
\subsection{ICPNe in the different fields} 
 
The observations of the different Virgo fields that we have used in
this study have slightly different filters, survey areas, and [OIII]
limiting magnitudes.  Depending on the cluster dynamics, the filter
bandpass may cut out a few ICPNe; the most likely field for this is
the FCJ field\footnote{For the mean redshift of the Virgo cluster and
a velocity dispersion $\sigma_{cluster}= 800$ kms$^{-1}$, the width of
the FCJ on-band filter covers the velocities in the -1.5
$\sigma_{cluster}$ to + 1.8 $\sigma_{cluster}$.}.  Before we
compare the number density of ICPNe in these fields and derive the
average amount of intracluster light in the Virgo cluster core, we
should establish a sample of ICPNe with the same flux limit. To this
end we multiply the number of ICPNe for each field by the factor:
\begin{equation} 
\Delta \equiv \int_{M^{*}}^{M^{*}+1} N(m)dm/\int_{m^{*}}^{m_{\rm lim}}N(m) dm, 
\end{equation} 
where $N(m)$ is given by eq.~\ref{PNLF}, $M^{*}$ and $m^{*}$ denote
the absolute and the apparent magnitude of its bright cutoff, and
$m_{\rm lim}$ is the [OIII] limiting magnitude in each field. In
scaling the various field samples by this factor, we obtain the
respective number of ICPNe down to a limiting absolute magnitude
$M^{*}+1$. The number of emission line candidates in the LPC field was
not scaled, because we saw in Section~\ref{lyalpha} that they are all
compatible with being Ly$\alpha$ background galaxies. For this field
we give an upper limit for the amount of the ICL, assuming that the
number of ICPNe in the field is $<1$. For the Core field, we could not
fit the IC PNLF, because it is contaminated by spill-over objects for
all magnitudes.  Thus, to compute $\Delta$ for this field we have used
the PNLF of M87 (Ciardullo et al.\ 1998).  We have also checked that
both spill-over and Ly$\alpha$ contaminant do not affect the brightest
bins of the IC PNLF of RCN1, so that the LF fitted in Paper I is still
valid. We used this LF in order to scale the number of ICPNe in the
RCN1 field down to $M^{*}+1$. The scaling factors, final numbers, and
resulting surface densities of ICPNe for the different fields are given
in Table~\ref{npne}.

\subsection{From the ICPNe surface density to surface brightness of 
the ICL: the $\alpha$ parameter}\label{alp} 
 
In principle, determining the amount of ICL from the observed numbers
of ICPNe is straightforward.  From M\'endez et al. (1993), if
$\dot{\epsilon}$ is the specific PN formation rate, in PN yr$^{-1}$
$L_\odot^{-1}$, $L_T$ is the total bolometric luminosity of a sampled
population and $t_{PN}$ is the lifetime of a PN, which we take as
25,000 yr, then the corresponding number of PNe, $n_{PN}$, is 
\begin{equation}
n_{PN} = \dot{\epsilon} L_{T} t_{PN}.
\end{equation}
Theories of stellar evolution predicts that the specific PN formation
rate should be $\sim 2 \times 10^{-11}$~stars yr$^{-1} L_\odot^{-1}$,
nearly independent of population age or initial mass function (Renzini
\& Buzzoni 1986).  Every stellar system should then have $n_{PN} =
\alpha \times L_{T} = 50 \times 10^{-8}$~PN~$L_\odot^{-1} \times
L_{T}$. If the PNLF of eq.~2 is valid 8 magnitudes down from the
cut-off, one can determine the fraction of PNe within 2.5 magnitudes
of the cut-off, and thus define $\alpha_{2.5}$ as the number of PNe
within 2.5 magnitudes of $M^*$ associated with a stellar population of
total luminosity $L_{T}$. Approximately 1 out of 10 of these PNe are
within 2.5 mags of $M^*$, and following from the above assumptions,
most stellar populations should have $\alpha_{2.5}
\sim 50\times 10^{-9}$~PNe~$L_\odot^{-1}$ (Feldmeier et al.\ 2004). 
The observed number of ICPNe can then be used to infer the total
luminosity of the parent stellar population.

However, as first noticed by Peimbert (1990), observations of PN
samples in galaxies show that $\alpha_{2.5}$ varies strongly as a
function of color. Hui et al.\ (1993) found that $\alpha_{2.5}$
decreases by a factor of 7 from the value of $\approx 50 \times
10^{-9}\,\mbox{PN}\, L_{\odot}^{-1}$ measured in dwarf ellipticals
like NGC 205 (Burstein et al.\ 1987) to $\approx 7 \times
10^{-9}\,\mbox{PN}\,L_{\odot}^{-1}$ observed in Virgo giant elliptical
galaxies (Jacoby et al.\ 1990).

The amount of ICL depends directly on the adopted value of $\alpha$,
which is thus not very well constrained and is a function of the
$(B-V)$ color of the parent stellar population, currently unknown for
the Virgo ICL. To take this uncertainty into account in our estimates
for the IC luminosity in the different fields, we consider three
plausible values for $\alpha$, which are i) the value appropriate for
an evolved population like that of the M31 bulge (Ciardullo et al.\
1989a); ii) the value determined by Durrell et al.\ (2002) for the
intracluster red-giant branch (RGB) stars observed with HST; and iii)
the $\alpha$ values determined from the Hui et al.\ (1993) empirical
relation. In this case the $(B-V)$ of the parent stellar population is
determined from the average colors of the Virgo galaxies near the
field position.

The scaling adopted in our work ensures that we have taken into
account all {\it bona-fide} ICPNe within 1 mag of $M^{*}$. We then
shall use the correspondent value $\alpha_{1.0}$ to infer the amount
of IC luminosities in our fields; from eq.\ref{PNLF},
$\alpha_{1.0}/\alpha_{2.5} \approx 0.24$.  Furthermore we adopt the
conversion from bolometric to B-band luminosity as in Ciardullo et
al.\ (1989a). Table~\ref{tabicl} gives the resulting values for the B-band
luminosity and surface brightness in our fields, in each case for the
three values of $\alpha$ (i-iii). The interval given by the lowest and
the largest value of the luminosity (surface brightness) thus
determined from the number of ICPNe in each field, is our {\it best
estimated range} for the IC luminosity (surface brightness) in that
field.

\subsection{Field-to-field variation and the dependence with distance
from the cluster center}  

We consider first the field-to-field variations in the number of ICPN
candidates as reported in Table~\ref{npne}.  Figure~\ref{result} shows
the ICPN number densities in our fields, with their Poisson
errors. The over-density observed at the FCJ field
position is significant at the 2-$3\sigma$ level with respect to the
number densities observed in Core and SUB.  The low density of ICPNe
in the LPC field is significant at the $ \sim 4\sigma$ level with
respect to the number densities observed in FCJ, Core and SUB. The low
upper limit to the number of ICPNe in LPC confirms earlier findings by
Kudritzki et al.\ (2000) for an adjacent field, centred at
$\alpha(J2000) = 12:26:32.1$, $\delta(J2000) = +12:14:39$.

The number density plot also shows no clear trend with distance from
the cluster center at M87, except that the value in the innermost FCJ
field is high.  However, the spectroscopic results of Arnaboldi et
al.\ (2004) have shown that 12/15 PNe in this field have a low
velocity dispersion of $250$ km/s, i.e., in fact belong to the outer
halo of M87, which thus extends to at least 65 kpc radius. In the SUB
field, 8/13 PNe belong to the similarly cold, extended halo of M84,
while the remaining PNe are observed at velocities that are close to
the systemic velocities of M86 and NGC 4388, the two other large
galaxies in or near this field. It is possible that in a cluster as
young and unrelaxed as Virgo, a substantial fraction of the ICL is
still bound to the extended halos of galaxies, whereas in denser and
older clusters these halos might already have been stripped. If so, it
is not inappropriate to already count the luminosity in these halos as
part of the ICL. However, in Fig.~\ref{result} we also give the plot
of the PN number density with radius for the case when the PNe in the
outer halos of M87 and M84 are removed from the FCJ and SUB
samples. In this case, the resulting number density is even more
nearly flat with radius, but there are still significant
field-to-field variations; particularly the remaining number density
in SUB and that in LPC are low.

We next consider the luminosity densities and surface brightnesses of
the ICL in our fields and their dependence on the distance from the
cluster center. This introduces the additional uncertainty of the
$\alpha$ parameter; when the range of $\alpha$ from Section
\ref{alp} is taken into account, the total uncertainties in the
luminosity density increase and differences between the various fields
become less significant. Data are given in Table~\ref{tabicl}. The
lower panel of Fig.~\ref{result} shows the radial variation of the
surface brightness as a function of distance from M87. In the FCJ and
SUB fields, we again show two points, respectively, one derived from
all PNe, and one based on only the PNe that are not kinematically
associated with the M87 or M84 halo.  Again, there is thus no clear
radial trend with distance: the ICL surface brightness in Virgo is
consistent with a constant value, with the exception of the low value
in the LPC field. As representative values for the surface luminosity
density and the surface brightness in the Virgo core region we take
the average for our 4 fields, including the M87 and M84 halo PNe,
and using for each field the respective average value of
$\alpha$ in Table~\ref{tabicl}: $\Sigma_B = 2.7 \times 10^{6}$
L$_{B\odot}$ arcmin$^{-2}$, and $\bar{\mu}_{B}=29.0$ mag
arcsec$^{-2}$. Note, however, the large $\mbox{rms} = 2.1 \times
10^{6}$ L$_{B\odot}$ arcmin$^{-2}$, due to the field-to-field
variations and the adopted range in $\alpha$.

Ferguson et al.\ (1998) and Durrell et al.\ (2002) measured the ICL in
two HST fields near M87 by counting excess intracluster RGB stars. One
can estimate the surface brightness of the underlying stellar
population, given assumptions on the metallicity, age, initial mass
function (IMF) and average distance for this stellar population.  From
the total flux associated with the 630 excess sources detected in
their HST field, Ferguson et al.\ (1998) computed that the surface
brightness and surface luminosity density of the ICL in their field
45.8 arcmin NE of M87 are $31.11$ mag arcsec$^{-2}$, and 0.42$\times
10^{6}$ L$_{B\odot}$ arcmin$^{-2}$, respectively.  Durrell et al.\
(2002) used the same technique in a field 42.3 arcmin NW of M87, and
detected $\sim 50\%$ greater number of stars than Ferguson et al.\
(1998). From their best fit models to the luminosity function, Durrell
et al. obtain $\mu_B = 30.07 $ and $29.677$ for the NE and NW HST
fields, respectively. We have over-plotted these values in
Figure~\ref{result}. It can be seen that they are in good agreement
with our inferred surface brightness, and still no clear radial trend
is evident. Averaging our results and the RGB data, the mean surface
luminosity density and surface brightness of the ICL in the combined
fields in the Virgo core region (the RCN1 field not included) become
$2.2\times 10^{6}$ L$_{B\odot}$ arcmin$^{-2}$ and 29.3 mag
arcsec$^{-2}$, respectively.

\subsection{Luminosity fraction of the ICL in the Virgo core} 

Surface luminosity densities associated with both ICPNe and RGB depend
for a large part on the assumed metallicity, age and IMF for the
parent stellar population in the LOS. The fact that our best estimates
are given as intervals in luminosities reflects a limitation in our
knowledge of the stellar populations and of post-AGB stellar
evolution, rather than uncertainties in the ICPN numbers from
photometric catalogs, which are now well understood.

When we compare the luminosity of the ICL at the positions of our
fields with the luminosity from the Virgo galaxies we add in further
uncertainties, because the luminosities of nearby Virgo galaxies depend
very much on the location and field size surveyed in the Virgo
cluster. We therefore consider the reported intervals in surface
brightness as our primary result, while the relative fractions of ICL
light with respect to Virgo galaxy light are evaluated for comparison
with previous ICPN works only, and we consider them as more uncertain.

Our approach is to assume that the ICL is related to local galaxies in
each field, based on the spectroscopic results of Arnaboldi et al.\
(2004) which show the Virgo cluster to be very inhomogeneous and still
dynamically young. We thus compare the luminosity of the ICL with that
from the Virgo galaxies located in the field.  We have obtained from
the DSS archive the images centered on our fields with a radius of
120$'$, and convolved each image with a Gaussian with $\sigma$ equal
to the radius of our corresponding images. Then, we determined the
total luminosity in an area equal to the surveyed area of each
field. We thus obtained 1.67 $\times 10^{10} L_{B\odot}$, 1.40 $\times
10^{10} L_{B\odot}$, 1.76 $\times 10^{10} L_{B\odot}$ and 4.41 $\times
10^{9} L_{B\odot}$ for the Core, FCJ , LPC and RCN1 fields,
respectively. The galaxy light in the SUB field is dominated by the
three bright galaxies in this field, M86, M84 and NGC4388. Their
combined luminosity is 7.2 $\times 10^{10} L_{B\odot}$ (see Paper
II). The contribution of the ICL to the total light in each field
based on this local comparison is given in Table~\ref{tabicl}; the
mean fraction for our 4 fields in the Virgo core is 5\%.  However,
there are significant field-to-field variations. The fraction of ICL
versus total light ranges from $\simeq 8\%$ in the for Core and FCJ
fields, to less than 1$\%$ in the LPC field, which in its low ICL
fraction is similar to low density environments (Castro-Rodriguez et
al. 2003).

\subsection{Discussion} 
 
The data presented here constitute a sizeable sample of ICPNe in the
Virgo core region, constructed homogeneously and according to rigorous
selection criteria. From the study of four wide-fields in the Virgo
core we obtain a mean surface luminosity density of
2.7$\times 10^{6}$ L$_{B \odot}$ arcmin$^{-2}$, $\mbox{rms} = 2.1
\times 10^{6}$ L$_{B \odot}$ arcmin$^{-2}$, and a mean surface
brightness of $\mu_{B}=29.0$ mag arcsec$^{-2}$. Often these numbers
are translated into fractions of ICL with respect to light in galaxies
at the cluster positions, and our best estimate in the Virgo core is
$\sim 5\%$.

We can compare the amount of ICL in the Virgo cluster core region
with that measured in the RCN1 field (see Paper I).  The surface
luminosity density and surface brightness intervals evaluated for this
field are similar to those for the FCJ. The RCN1 field contains no
bright galaxies, however, so these values translate into a
significantly higher fraction of ICL with respect to Virgo galaxy
light, about $\simeq 34\%$.  If we include the RCN1
measurements in our average, the mean surface luminosity becomes
2.7$\times 10^{6}$ L$_{B \odot}$ arcmin$^{-2}$, $\mbox{rms} = 1.9
\times 10^{6}$ L$_{B \odot}$ arcmin$^{-2}$ and the mean surface
brightness $\mu_{B}=29.1$ mag arcsec$^{-2}$. The average fraction of
ICL with respect to Virgo cluster galaxy light then becomes $\sim
10\%$.

Previous works on ICL using ICPN tracers have claimed larger average
fractions for ICL in the Virgo core ($15.8\%$, Feldmeier et al.\ 2004;
$15\%$ Durrell et al. 2002), despite similar inferred surface
brightnesses for the ICL as shown in Figure~\ref{result}. These higher
fractions reflect only a difference in the normalisation relative to
the Virgo galaxy light. Furthermore, the Feldmeier et al. (2004)
fields in the Virgo subcluster A are closer on average to M87 ($\sim
25'$) than our fields; thus, they may be more dominated by the M87
halo, as indicated by the spectroscopic follow-up of the ICPN sample
in the FCJ field (Arnaboldi et al.\ 2004). These radial velocity
measurements indicate the presence of a stellar population at
equilibrium in the M87 halo, and only $3/15 = 1/5$ of all PN
candidates in the FCJ field are therefore ``truly'' intracluster. The
spectroscopic follow-up of Arnaboldi et al.\ (2004) has confirmed the
expected fraction of ICPN candidates in the other fields, therefore
supporting our determinations from the PNe observations.
 
The field-to-field variations in the measured number density of ICPNe 
and the high value in the outer RCN1 field indicate that the ICL is 
still not relaxed in the cluster potential. There has not been enough 
time for phase-mixing to erase these variations. This poses a strong 
constraint on the age and origin of the ICL in the Virgo cluster core 
region, which must have been brought into this location not much more 
than a few dynamical times ago, where (Binney \& Tremaine 1987) 
\begin{equation} 
t_{dyn}=\sqrt{\frac{3\pi}{16 G \rho}}, 
\end{equation} 
and $\rho$ is the mean density enclosed within the field distance from
the center of the cluster. We have computed $t_{dyn}$ using the mass
distribution given by Nulsen \& Bohringer (1995) for the central
regions of the Virgo cluster. The dynamical times at the location of
our fields are from $\sim 2\times 10^{8}$ yr for the FCJ field to
$\sim 8\times 10^{8}$ yr for the LPC and SUB fields.  The FCJ field has
the shorter dynamical time because it is located close to M87
($\approx 65$ kpc). From these estimated dynamical times,
phase-mixing to erase the field-to-field variations in the ICL light
would take a few Gyr. The observed field-to-field variations thus
suggest that the Virgo cluster is dynamically young, confirming the
similar inference from the inhomogeneous radial velocity distributions
by Arnaboldi et al.\ (2004).
 
Recent cosmological simulations have measured the amount of diffuse 
light in simulated galaxy clusters (Murante et al. 2004; Sommer-Larsen 
et al. 2004). They conclude that the mass of the diffuse light varies 
with the mass of the cluster, being larger for more massive clusters. 
Murante et al.\ (2004) studied the diffuse light for clusters with 
masses in the range $10^{14}-10^{15} M_{\odot}$. They found for 
low-mass clusters like Virgo ($\approx 10^{14} M_{\odot}$) that the 
surface mass density of the diffuse light in the innermost regions of 
the clusters is $\approx 10^{7} h M_{\odot}$ kpc$^{-2}$,
which is in agreement with our measured mean surface 
luminosity density of the ICL in the Virgo core of 2.7$\times 10^{6}$ 
L$_{B\odot}$ arcmin$^{-2}$, for a mass-to-light ratio of $\sim 5$, as
expected for an evolved stellar population. 
  
\section{Conclusions} 
 
Four pointings around the core region of the Virgo cluster have been 
analyzed to determine the amount of intracluster light (ICL) via 
intracluster planetary nebulae (ICPNe). The photometric ICPN 
candidates, for three of the fields, were selected according to the 
on-off band technique using [OIII]$\lambda5007$\AA, and for the 
fourth field by the emission in [OIII] and H$\alpha$. We have 
carefully studied the contribution of different contaminants to the 
photometric samples of ICPNe, using simulations on our images. In this 
way we obtain a well-understood, magnitude-limited sample of ICPN 
candidates. 

In the SUB field, spectroscopic follow-up has provided kinematic
evidence that the so-called 'over-luminous' PNe in the halo of M84 are
dynamically associated with this galaxy. They must therefore be
intrinsically brighter than the normal elliptical galaxy PN population
in M84, presumably due to a younger age or different metallicity.
 
Based on our new sample of ICPNe,  the mean surface brightness and
surface luminosity density of the ICL in our 4 fields are
$\mu_{B}=29.0$ mag arcsec$^{-2}$ and 2.7$\times 10^{6}$ L$_{B
\odot}$ arcmin$^{-2}$, respectively. These values are in good
agreement with the corresponding values obtained from excess red giant
counts in two HST images in the Virgo core. There is no trend evident
with distance from M87. When the fraction of ICL is computed for our
four fields in the Virgo core, it amounts to 5$\%$ of the total galaxy
light.
 
However, the diffuse stellar population in Virgo is inhomogeneous on
scales of $30'-90'$: we observe substantial field-to-field variations
in the number density of PNe and the inferred amount of intracluster
light, with some empty fields, some fields dominated by extended Virgo
galaxy halos, and some fields dominated by the true intracluster
component.

The field-to-field variations indicate that the ICL is not yet
dynamically mixed.  This imposes a constraint on the time of origin of
the ICL and the Virgo cluster itself. The lack of phase-mixing
suggests that both have formed in the last few Gyr, and that local
processes like galaxy interactions and harassment have played an
important role in this.  In a cluster as young and unrelaxed as Virgo,
a substantial fraction of the ICL may still be bound to the extended
halos of galaxies, whereas in denser and older clusters these halos
might already have been stripped.

\acknowledgments We wish to thank an anonymous referee for
insightful comments.  We thank ESO, Subaru and ING for the observing
time allocated to this project. We acknowledge financial support by
SNF grants 20-56888.99 and 200020-101766, and by INAF - Projects of
national interests, PI: MA. JALA acknowldeges funding by the Spanish DGES,
grant AYA2001-3939; NRN was supported by a Marie Curie fellowship.

 
 
 
{} 
 
\clearpage 
 
\begin{deluxetable}{ccccccccccc} 
\tabletypesize{\scriptsize}  
\tablecaption{Summary of field positions and filter characteristics}      
\tablewidth{0pt}      
\tablehead{     \colhead{Field}     & 
\colhead{$\alpha$(J2000)}      &      \colhead{$\delta$(J2000)}      & 
\multicolumn{2}{c}{[{\rm OIII}]  Filter}   &  &  \multicolumn{2}{c}{Off-band 
Filter} & C$^{a}$ & m$_{\rm lim}(5007)$ & $m_{\rm lim}({\rm off})$ \\  
\cline{4-5} \cline{7-8} \\ & \colhead{(hh:mm:ss)} 
&   \colhead{($^{o}$:$'$:$''$)}  &  \colhead{$\lambda_{c}$   (\AA)}  & 
\colhead{FWHM   (\AA)}   &   &   \colhead{$\lambda_{c}$   (\AA)}   & 
\colhead{FWHM (\AA)} & \\ }  
\startdata  
Core & 12:27:48 & +13:18:46 & 5023 & 80 & & 5395 & 894  &2.51 & 27.21 
& 24.75\\  
FCJ & 12:30:39 & +12:38:10 & 5027 & 44 & & 5300 & 267 &2.50 & 27.01 & 24.58\\  
LPC  & 12:25:32 & +12:14:39 & 5027 & 60 & & 4407 &  1022 &3.02 & 27.52 
& 25.40 \\   
SUB & 12:25:47  & +12:43:58 & 5021  & 74 &  & & &2.49 & 28.10 & \\  
RCN1 & 12:26:13 & +14:08:03 & 5023 & 80 & & 5395 & 894 &2.51 & 26.71&24.75 \\  
\enddata  
\tablenotetext{a}{C is  the transformation constant 
between AB  and 5007  magnitudes for the  narrow band  [OIII] filters: 
m(5007)=m(AB)+C.} 
\end{deluxetable}

\clearpage 
 
\begin{deluxetable}{ccccccc} 
\tabletypesize{\normalsize}   
\tablecaption{Number of ICPNe in surveyed fields}  
\tablewidth{0pt}  
\tablehead{ \colhead{Field} & $N_{ini}^{a}$ & $N_{mlim}^{b}$ & 
  $N_{so}^{c}$ & $N_{lost}^{d}$ & $N_{Ly\alpha}^{e}$ & $N_{final}^{f}$ \\ }  
\startdata  
Core & 117 & 77 & 45& 1 & 20/26  & 13  \\  
FCJ  & 36  & 20 & 4 & 2 & 2/4    & 16 \\  
LPC  & 14  & 14 & 2 & 0 & 22/16  & $<1$  \\  
SUB  & 36  & 36 & - & - &  -     & 36    \\ 
RCN1 & 75  & 55 & 16& 1 & 3/26   & 37/15 \\ 
\enddata 
\tablenotetext{a}{Total number of objects obtained from the CMDs with 
  our selection criteria as bona-fine emission line objects.}  
\tablenotetext{b}{Number of selected emission line objects down to the 
  [OIII] limiting magnitude of each field.}  
\tablenotetext{c}{Number of contaminants due to spill-over effect.}  
\tablenotetext{d}{Number of objects with EW$_{obs}>110$\AA\ lost due to 
  photometric errors.}  
\tablenotetext{e}{Number of contaminants due to background Ly$\alpha$ 
  galaxies.}  
\tablenotetext{f}{Final estimated number of ICPNe.}  
\end{deluxetable}

\clearpage 
 
\begin{deluxetable}{cccccccccc}
\tabletypesize{\scriptsize}
\tablecaption{Number of ICPNe within $M^*+1$ in surveyed fields\label{npne}}
\tablewidth{0pt}
\tablehead{ \colhead{Field} & $N_{mlim}^{a}$ & $\Delta^{b}$ &
  $N_{ICPNe}^{c}$ & Area (arcmin$^{2}$) & ICPNe/arcmin$^{2}$ \\ }
\startdata
Core& 13 & 1.28 & 17  & 943 & 0.018    \\
FCJ & 16 & 0.98 & 15  & 266 & 0.056    \\
LPC &  1 &      & 1   & 744 & 0.001    \\
SUB & 36 & 0.40 & 14  & 706 & 0.020    \\
RCN1 & 37/15 & 1.15   & 43/17 & 957 & 0.045/0.018
\enddata
\tablenotetext{a}{Number of estimated emission line objects brighter
than the [OIII] limiting magnitude in each field.}
\tablenotetext{b}{Scale factor}
\tablenotetext{c}{Final number of ICPNe}
\end{deluxetable}

\clearpage

\begin{deluxetable}{ccccc}
\tabletypesize{\scriptsize}
\tablecaption{Surface brightness of the ICL in the surveyed fields}
\tablewidth{0pt}
\tablehead{ \colhead{Field} & $\alpha_{1.0,B}$ & $L_{ICL}/arcmin^{2}$
  &  $\mu_{B}$ & $\% Local^{a}$ \\
 & ($\times 10^{-9}$) & ($\times 10^{6} L_{\odot,B}/arcmin^{2}$) &
  (mag/arcsec$^{2}$) & \\}
\startdata
Core & 9.40   & 1.91   & 29.5   &10  \\
     & 13.30  & 1.35   & 29.8   & 7  \\
     & 10.59  & 1.70   & 29.6   & 9  \\
FCJ  & 9.40   & 5.96   & 28.2   & 8  \\
     & 13.30  & 4.21   & 28.6   & 6  \\
     & 7.46   & 7.51   & 28.0   & 10 \\
LPC  & 9.40   & 0.11   & 32.6   &0.5 \\
     & 13.30  & 0.08   & 32.9   &0.3 \\
     & 25.44  & 0.04   & 33.7   &0.2 \\
SUB  & 9.40   & 2.13   & 29.4   & 2  \\
     & 13.30  & 1.50   & 29.7   & 1  \\
     & 3.46   & 5.78   & 28.3   & 5  \\
RCN1 & 9.40   & 4.79/1.91 & 28.5/29.5 & 51/29   \\
     & 13.30  & 3.38/1.35 & 28.9/29.9 & 42/23   \\
     & 15.00  & 3.00/1.20 & 29.0/30.0 & 39/21   \\
\enddata
\tablenotetext{a}{Contribution of the ICL to the total light in
galaxies at the local position of the field.\label{tabicl}}

\end{deluxetable}
 
\clearpage 
 
 
\begin{figure} 
\epsscale{1.0} 
\plotone{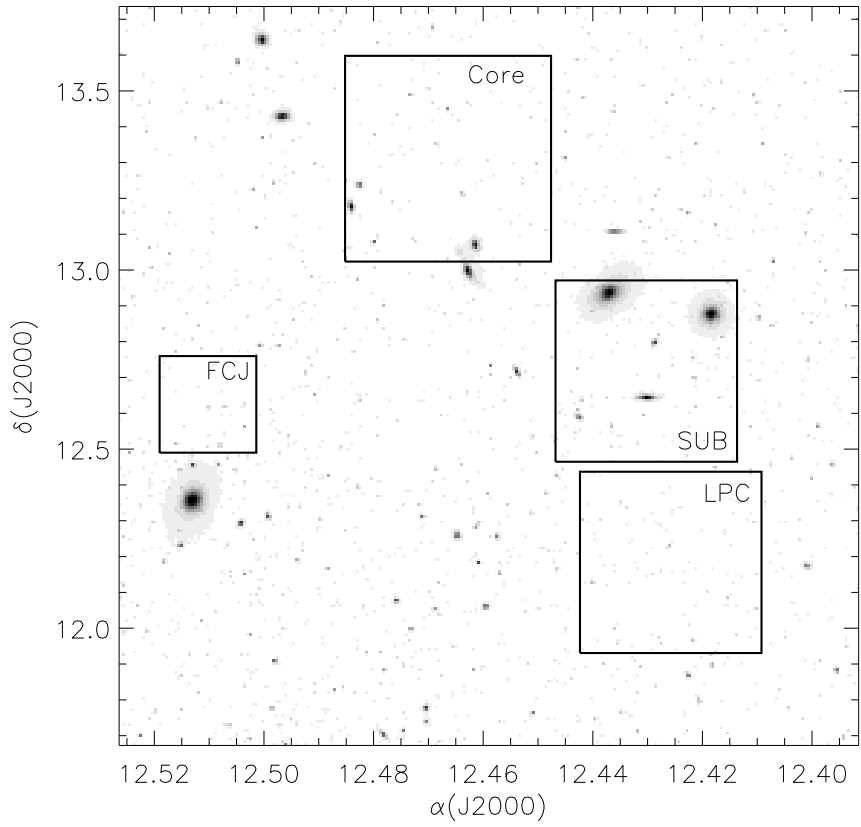} 
\caption{Virgo cluster core region with the positions of the fields studied 
in this work. The adopted center for Virgo cluster is M87. 
\label{fields}} 
\end{figure} 
 
\clearpage 
 
\begin{figure} 
\epsscale{0.6} 
\plotone{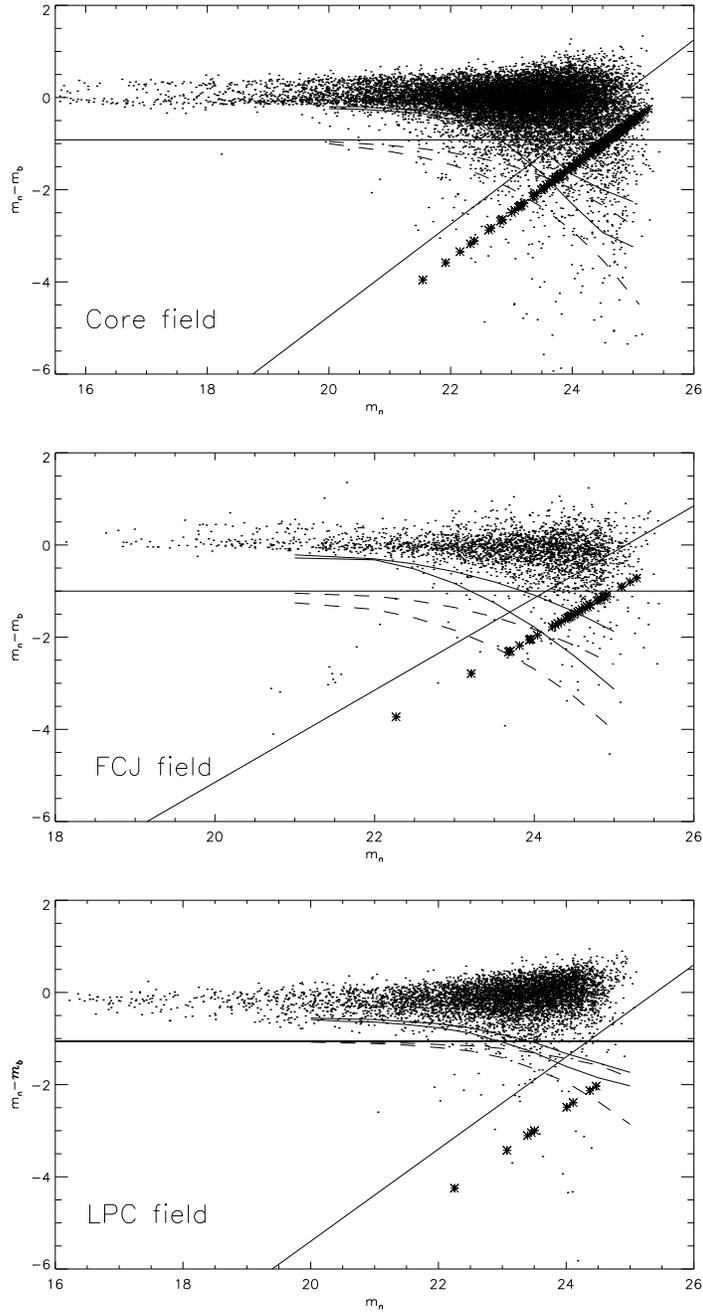} 
\caption{Color magnitude diagrams of the Core (top), FCJ (middle) and LPC 
  (bottom) fields. The horizontal full line represents objects with 
  observed EW=100\AA. The diagonal line shows the magnitude 
  corresponding to 1.0$\times \sigma$ above the sky in the off band 
  images. The full curved lines represent the 99$\%$ and 99.9$\%$ 
  lines for the distribution of modeled continuum objects. The dashed 
  curved lines represent the 84$\%$ and 97.5$\%$ lines for the 
  distribution of modeled objects with observed EW=110\AA. The 
  points are all detected objects by SExtractor. The asterisks 
  represent those objects with no broad band magnitude measured by 
  SExtractor. \label{fig2}} 
\end{figure} 
 
\clearpage  
 
\begin{figure} 
\epsscale{1.0} 
\plotone{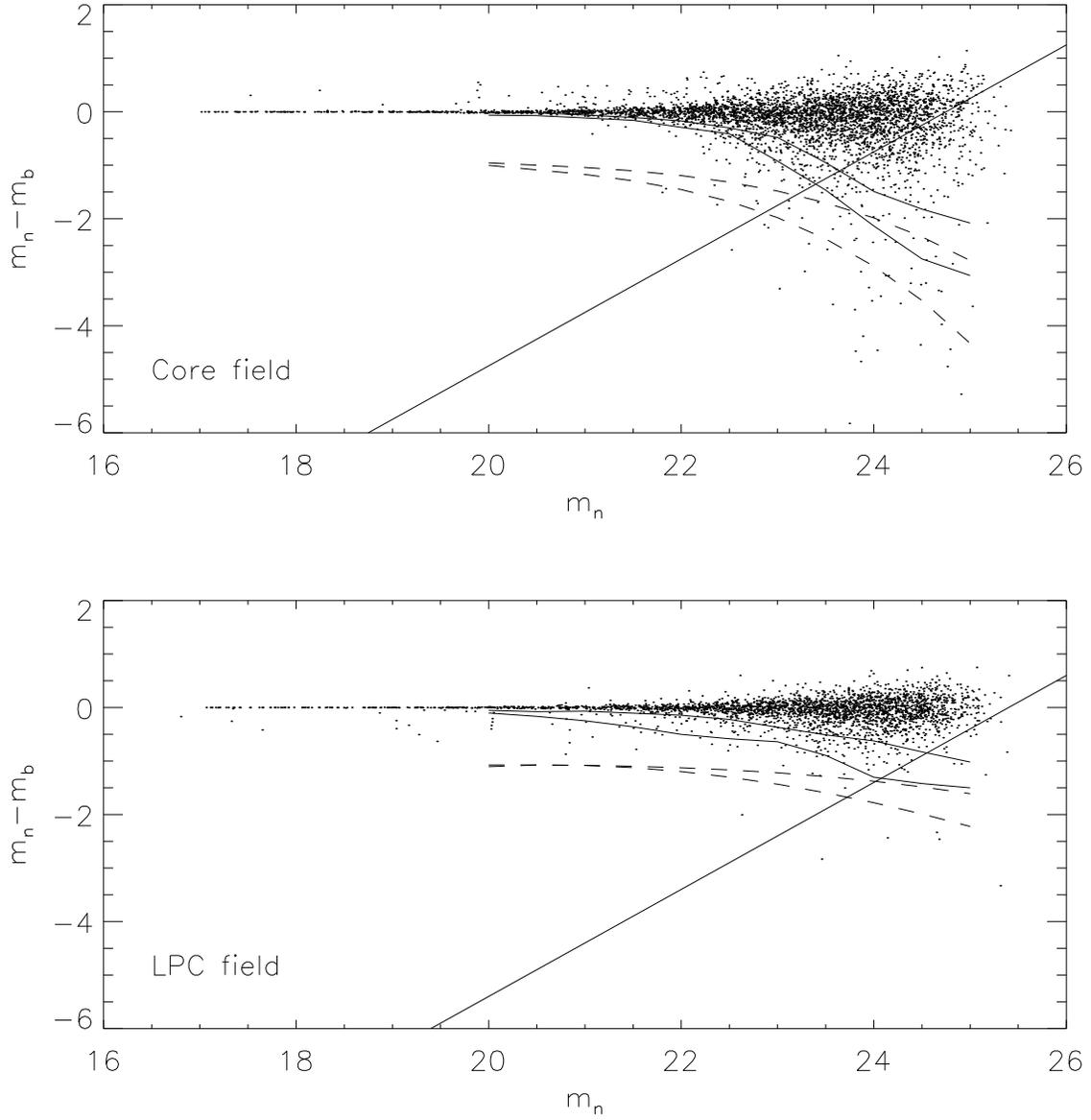} 
\caption{Color magnitude diagrams of simulated continuum objects in
the Core (top) and LPC (bottom) fields, to quantify the the spill-over
effect from faint stars. The dashed and full lines have the same
meaning as in Figure~\ref{fig2}. \label{fig3}} 
\end{figure} 
 
\clearpage  
 
\begin{figure} 
\epsscale{1.0} 
\plotone{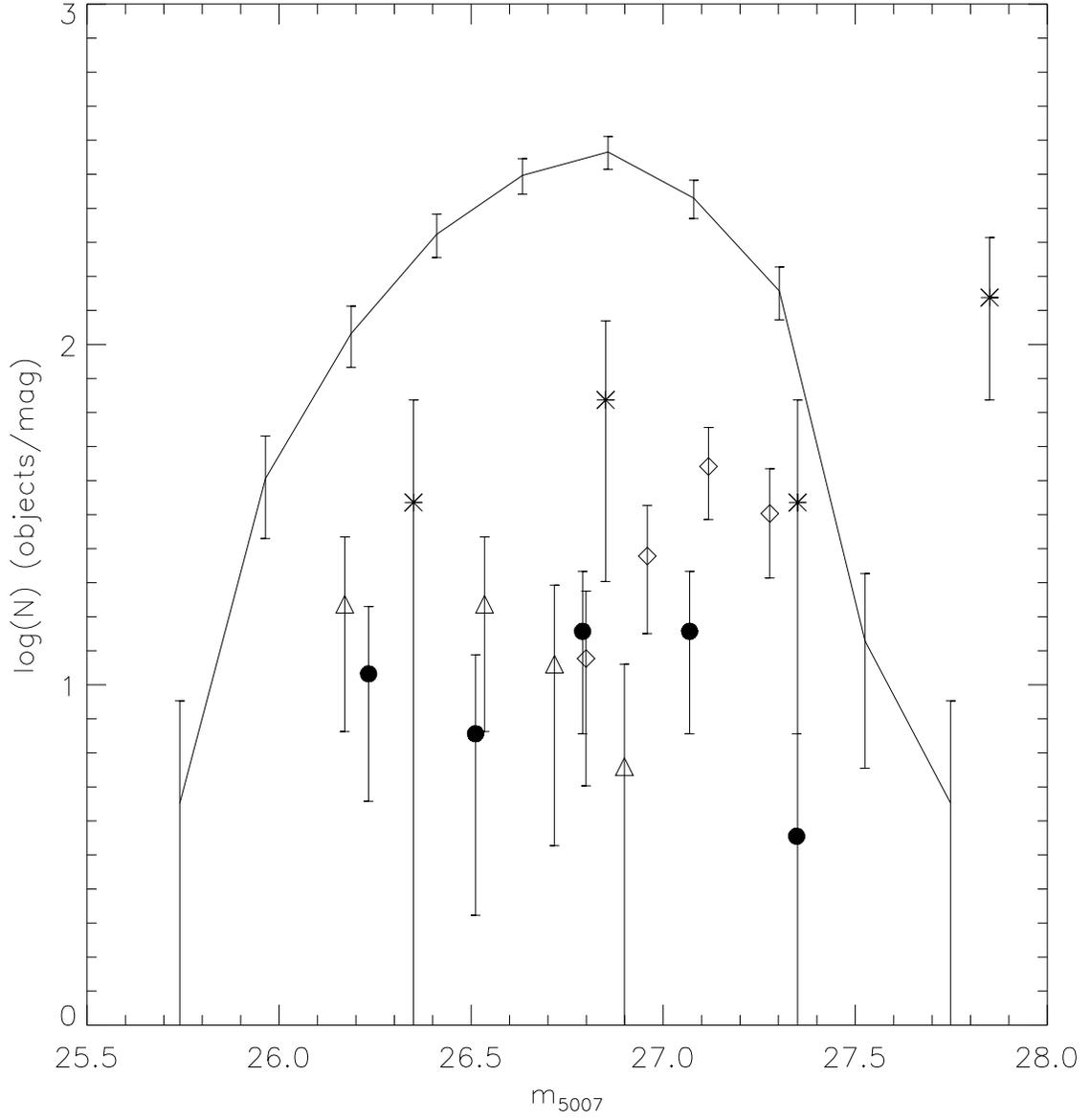} 
\caption{ICPN LF in the LPC field (full points) compared with the
Ly$\alpha$ LF from Kudritzki et al.\ (2000) (asterisks), Ciardullo
et al.\ (2002) (triangles), and our Leo field (diamonds). The LF of
the Ly$\alpha$ emitters were scaled to the effective surveyed volume
of the LPC field. The continuous line indicate the PNLF of M87, from
the inner $5'$ region.\label{fig4}}
\end{figure} 

\clearpage  
 
\begin{figure} 
\epsscale{1.0} 
\plotone{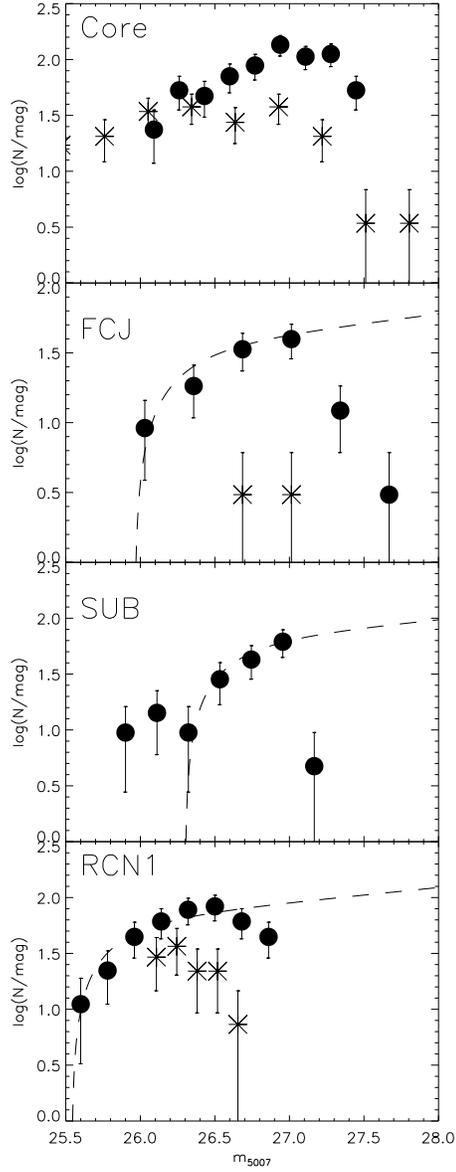} 
\caption{ LF of all ICPN photometric candidates (full points) in the
Core, FCJ, SUB and RCN1 fields. The asterisks represent the LF of the
simulated spill-over stars. The dashed-lines represent the fitted LFs.
\label{fig5}} 
\end{figure} 
 
\clearpage  
 
\begin{figure} 
\epsscale{1.0} 
\plotone{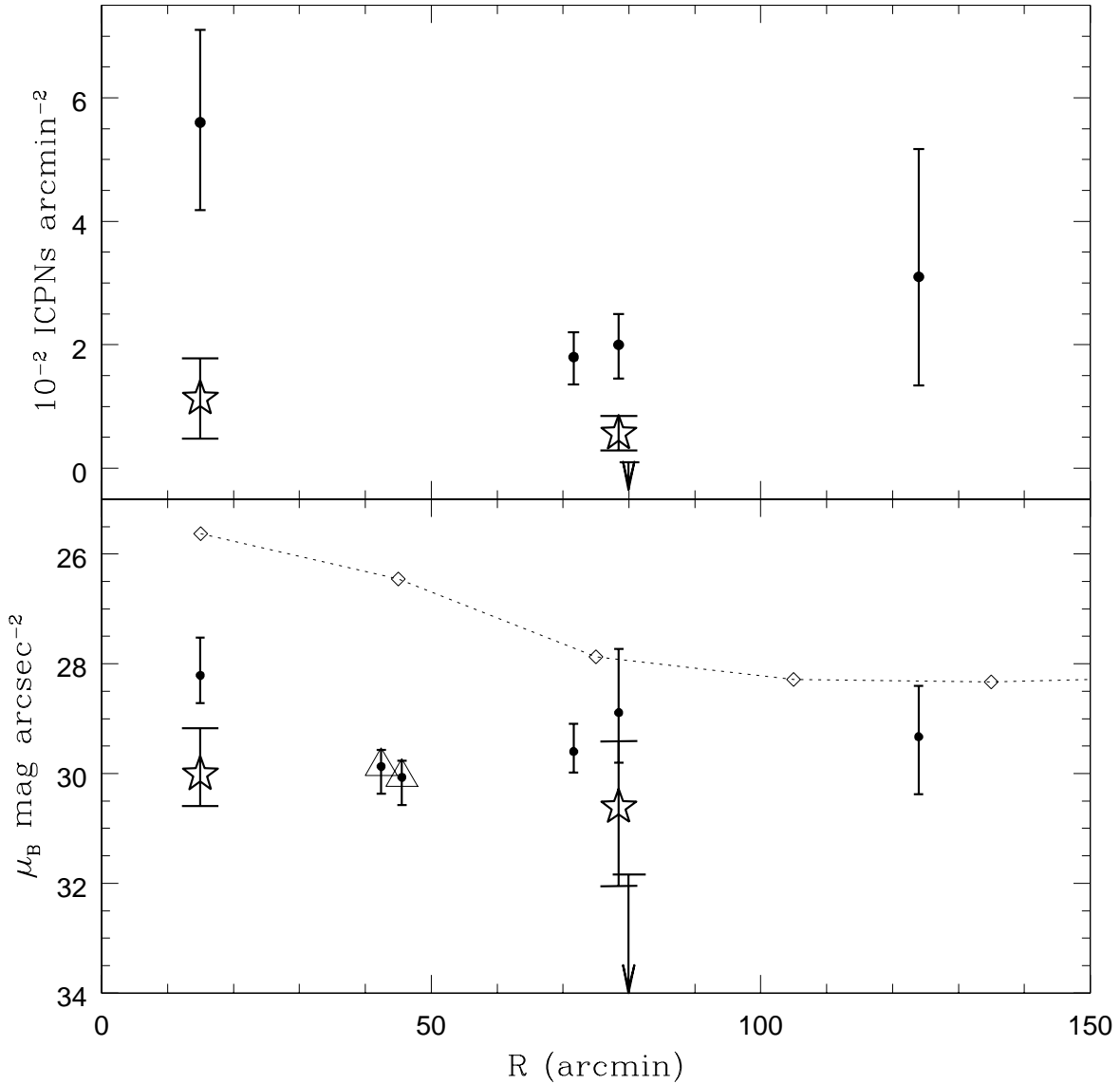} 
\caption{Number density of PNe (upper panel) and surface brightness
(lower panel) in our surveyed fields. In the upper panel, circles show
the measured number densities from Table \ref{npne}, and errorbars
denote the Poisson errors. For the LPC field our upper limit is
given. For the RCN1 field at the largest distance from M87, the
uncertainty from the correction for Ly$\alpha$ emitters is substantial
and is included in the errorbar. The large stars with Poisson error
bars show the number densities of PNe in FCJ and SUB fields not
including PNe bound to the halos of M87 and M84.  In the lower panel,
circles show the surface brightness inferred with the average value of
$\alpha$ in Table \ref{tabicl}, and the errorbars show the range of
values implied by the Poisson errors and the range of adopted $\alpha$
values. Triangles represent the measurements of the ICL from RGB
stars; errorbars indicate uncertainties in the metallicity, age and
distance of the parent population as discussed in Durrell et al.\
(2002).  The stars indicate the surface brightness associated with the
ICPNe in the FCJ and SUB fields, that are not associated to the M87 or
M84 halos but are free flying in the Virgo cluster potential, from
Arnaboldi et al.\ (2004). The dashed line and diamonds show the B band
luminosity of Virgo galaxies averaged in rings, from Binggeli et al.\
(1987). Distances are relative to M87. The ICL shows no trend with
cluster radius out to $150'$.
\label{result}} 
\end{figure} 
 

\begin{thebibliography}{} 

\bibitem[Alcal\'a et al.(2004)]{1053} Alcal\'a, J.~M., et al.\ 2004, \aap,
428, 339
 
\bibitem[Alcal\'a et al.(2002)]{1056} Alcal\'a, J.~M., et al.\ 2002, \procspie, 
4836, 406  

\bibitem[Arnaboldi et al. (2004)]{1059} Arnaboldi,  M.~et al.\  2004, \apj, 
 614, L33 

\bibitem[Arnaboldi  et al.(2003)]{1062} Arnaboldi,  M.~et al.\  2003, \aj, 
  125, 514 (Paper II) 

\bibitem[Arnaboldi  et al.(2002)]{1065} Arnaboldi,  M.~et al.\  2002, \aj, 
123, 760 (Paper I) 

\bibitem[Arnaboldi et  al.(1998)]{1068} Arnaboldi, M.~et  al.\ 1998, \apj, 
507, 759

\bibitem[Arnaboldi et  al.(1996)]{1071} Arnaboldi, M.~et  al.\ 1996, \apj, 
472, 145 
 
\bibitem[Bernstein  et al.(1995)]{1074}  Bernstein, G.~M.,  Nichol, R.~C., 
Tyson, J.~A., Ulmer, M.~P., \& Wittman, D.\ 1995, \aj, 110, 1507 

\bibitem[Bertin \& Arnouts(1996)]{1077} Bertin, E.~\&  
Arnouts, S.\ 1996, \aaps, 117, 393 

\bibitem[Binggeli, Tammann, \& Sandage(1987)]{1080} Binggeli, B., Tammann, 
G.~A., \& Sandage, A.\ 1987, \aj, 94, 251 

\bibitem[]{1083} Binney, J. J., \& Tremaine, S. 1987, Galactic Dynamics 
  (Princenton: Princeton Univ. Press) 
 
\bibitem[Burstein et al.(1987)]{1086} Burstein, D., Davies,  
R.~L., Dressler, A., Faber, S.~M., Stone, R.~P.~S., Lynden-Bell, D.,  
Terlevich, R.~J., \& Wegner, G.\ 1987, \apjs, 64, 601 

\bibitem[Castro-Rodr{\'{\i}}guez et al.(2003)]{1096}
Castro-Rodr{\'{\i}}guez, N., Aguerri, J.~A.~L., Arnaboldi, M.,
Gerhard, O., Freeman, K.~C., Napolitano, N.~R., \& Capaccioli, M.\
2003, \aap, 405, 803
 
\bibitem[Ciardullo et al. (2002a)]{1101} Ciardullo, R.,  
Feldmeier, J.~J., Krelove, K., Jacoby, G.~H., \& Gronwall, C.\ 2002a, \apj,  
566, 784  

\bibitem[Ciardullo et al. (2002b)]{1105} Ciardullo, R., Feldmeier, J.J.,
Jacoby, G.H., Kuzio de Naray, R., Laychak, M.B., Durrell, P.R.\ 2002b, \apj,  
577, 31

\bibitem[Ciardullo, Jacoby, Feldmeier, \&  
Bartlett(1998)]{a} Ciardullo, R., Jacoby, G.~H.,  
Feldmeier, J.~J., \& Bartlett, R.~E.\ 1998, \apj, 492, 62  
  
\bibitem[Ciardullo, Jacoby, Ford, \& Neill(1989)]{1113}  
Ciardullo, R., Jacoby, G.~H., Ford, H.~C., \& Neill, J.~D.\ 1989a, \apj,  
339, 53  
 
\bibitem[Ciardullo, Jacoby, \& Ford(1989a)]{1117} Ciardullo,  
R., Jacoby, G.~H., \& Ford, H.~C.\ 1989b, \apj, 344, 715  


\bibitem[Colless, Ellis, Taylor, \& Hook(1990)]{1121}  
Colless, M., Ellis, R.~S., Taylor, K., \& Hook, R.~N.\ 1990, 
\mnras, 244, 408  
 
\bibitem[Dopita, Jacoby, \& Vassiliadis(1992)]{1128} Dopita,  
M.~A., Jacoby, G.~H., \& Vassiliadis, E.\ 1992, \apj, 389, 27 
 
\bibitem[Dubinski(1998)]{1131} Dubinski, J.\ 1998, \apj, 502, 141 
 
\bibitem[Durrell et al.(2002)]{1133} Durrell, P.~R.,  
Ciardullo, R., Feldmeier, J.~J., Jacoby, G.~H., \& Sigurdsson, S.\ 2002,  
\apj, 570, 119  
 
\bibitem[Feldmeier, Ciardullo, Jacoby, \& Durrell(2004)]{1137} Feldmeier, 
J.~J., Ciardullo, R., Jacoby,  G.~H., \& Durrell, P.~R.\ 2004, \apj,
615, 196. 
 
 
\bibitem[Feldmeier, Ciardullo, Jacoby, \& Durrell(2003a)]{1142} Feldmeier, 
J.~J., Ciardullo, R., Jacoby,  G.~H., \& Durrell, P.~R.\ 2003a, \apjs, 
145, 65 
 
\bibitem[Feldmeier, Ciardullo, Jacoby, \& Durrell(2003b)]{1146} Feldmeier, 
J.J., Ciardullo, R.,  Jacoby, G., \& Durrell, P.\  2003b, in IAU Symposium 
217, Recycling Intergalactic and Interstellar Matter, in press 
 
\bibitem[Feldmeier,  Ciardullo, \&  Jacoby(1998)]{1150}  Feldmeier, J.~J., 
Ciardullo, R., \& Jacoby, G.~H.\ 1998, \apj, 503, 109 
 
\bibitem[Ferguson,  Tanvir, \&  von  Hippel(1998)]{1153} Ferguson,  H.~C., 
Tanvir, N.~R., \& von Hippel, T.\ 1998, \nat, 391, 461 
 
\bibitem[Freeman  et  al.(2000)]{1156} Freeman,  K.~C.~et  al.\ 2000,  in ASP 
Conf.~Ser.~197: Dynamics  of Galaxies: from the Early  Universe to the 
Present, 389 
 
\bibitem[Gal-Yam, Maoz, Guhathakurta, \&  
Filippenko(2003)]{b} Gal-Yam, A., Maoz, D., Guhathakurta,  
P., \& Filippenko, A.~V.\ 2003, \aj, 125, 1087  
 
\bibitem[Gonzalez,   Zabludoff,    Zaritsky,   \&   Dalcanton(2000)]{1164} 
Gonzalez, A.~H., Zabludoff, A.~I.,  Zaritsky, D., \& Dalcanton, J.~J.\ 
2000, \apj, 536, 561 
 
\bibitem[Gregg  \& West(1998)]{1168}  Gregg, M.~D.~\&  West,  M.~J.\ 1998, 
\nat, 396, 549 
 
\bibitem[Hammer et al.(1997)]{1171} Hammer, F., et al.\  
1997, \apj, 481, 49  
 
\bibitem[Hogg, Cohen, Blandford, \& Pahre(1998)]{1174} Hogg, D.~W., Cohen, 
J.~G., Blandford, R., \& Pahre, M.~A.\ 1998, \apj, 504, 622 
 
\bibitem[Hui, Ford, Ciardullo, \& Jacoby(1993)]{1177} Hui,  
X., Ford, H.~C., Ciardullo, R., \& Jacoby, G.~H.\ 1993, \apj, 414, 463  
 
\bibitem[Jacoby, Ciardullo, \& Walker(1990)]{1180} Jacoby,  
G.~H., Ciardullo, R., \& Walker, A.~R.\ 1990, \apj, 365, 471 
 
\bibitem[Jacoby(1989)]{1183} Jacoby, G.~H.\ 1989, \apj, 339, 39 
 
\bibitem[Kudritzki et al.(2000)]{1185} Kudritzki, R.-P., et al.\ 2000, 
  \apj, 536, 19  
 
\bibitem[]{1188} Marigo, P., Girardi, L., Weiss, A., Groenewegen, 
  M. A. T., \& Chiosi, C., 2004, \aap, 423, 995

\bibitem[Malumuth \& Richstone(1984)]{1191} Malumuth,  
E.~M.~\& Richstone, D.~O.\ 1984, \apj, 276, 413 

\bibitem[Mendez et al. (1993)]{1194} Mendez, R. H., Kudritzki, R. P.,
Ciardullo, R., \& Jacoby, G. H.\ 1993, \aap, 275, 534
 
\bibitem[Merritt(1983)]{1197} Merritt, D.\ 1983, \apj, 264, 24  
 
\bibitem[Merritt(1984)]{1199} Merritt, D.\ 1984, \apj, 276, 26  
 
\bibitem[Miller(1983)]{1201} Miller, G.~E.\ 1983, \apj, 268, 495  
 
\bibitem[Moore  et  al.(1996)]{1203}  Moore,   B.,  Katz,  N.,  Lake,  G., 
Dressler, A., \& Oemler, A.\ 1996, \nat, 379, 613 
 
\bibitem[]{1206} Murante, G. et al., 2004, ApJ, 607, L83 
 
\bibitem[Napolitano  et al.(2003)]{1208}  Napolitano, N.~R.~et  al.\ 2003, 
\apj, 594, 172 
 
\bibitem[Neilsen \& Tsvetanov (2000)]{1211} Neilsen, E.H., \& Tsvetanov,
Z.I.\ 2000, \apj, 536, 255 

\bibitem[Nulsen \& Bohringer(1995)]{1214} Nulsen,  
P.~E.~J.~\& Bohringer, H.\ 1995, \mnras, 274, 1093 
 
\bibitem[Okamura et al.(2002)]{1217} Okamura, S., et al.\  
2002, \pasj, 54, 883  
 
\bibitem[Oemler(1973)]{1220} Oemler, A.\ 1973, \apj, 180, 11 

\bibitem[Peimbert (1990)]{1222} Peimbert, M.\ 1990, Rep. Prog. Phys., 53, 1559 

\bibitem[Renzini \& Buzzoni(1986)]{1224} Renzini, A.~\&  
Buzzoni, A.\ 1986, ASSL Vol.~122: Spectral Evolution of Galaxies, 195  
 
\bibitem[Richstone \&  Malumuth(1983)]{1230} Richstone, D.~O.~\& Malumuth, 
E.~M.\ 1983, \apj, 268, 30 
 
\bibitem[]{1233} Sommer-Larsen, J., Romeo, A. D. \& Portinari, L., 2004, 
  MNRAS, submitted (astro-ph/0403282) 
 
\bibitem[Teplitz et al.(2000)]{1236} Teplitz, H.~I., et al.\  
2000, \apj, 542, 18 
 
\bibitem[Theuns \& Warren(1997)]{1239}  Theuns, T.~\& Warren, S.~J.\ 1997, 
\mnras, 284, L11 
 
\bibitem[Thuan  \&  Kormendy(1977)]{1242}  Thuan, T.~X.~\&  Kormendy,  J.\ 
1977, \pasp, 89, 466 
 
\bibitem[West \& Blakeslee(2000)]{1245} West, M.~J.~\&  
Blakeslee, J.~P.\ 2000, \apjl, 543, 27  

\bibitem[Willman et al. (2004)]{1248} Wilman, B., Governato, F., Wadsley,
J., \& Quinn, T. 2004, \mnras, 355, 159
 
\bibitem[Zibetti, et al.\  (2005)]{1254} Zibetti, S., White, S.~D.~M.,
Schneider, D.M., \& Brinkmann, J., 2005, MNRAS, in press (astro-ph/0501194)
 
\bibitem[Zwicky(1951)]{1258} Zwicky, F.\ 1951, \pasp, 63, 61 
 
 
\end{thebibliography}
\end{document}